\documentclass{osa-article}

\journal{oe}

\articletype{Research Article}
\usepackage{caption}
\usepackage{subcaption}
\usepackage{lineno} 
\usepackage{amsmath}

\begin{document}

\title{Performance and limitations of dual-comb based ranging systems}

\author{Bruno Martin \authormark{1,2}\authormark{*}, Patrick Feneyrou\authormark{3}, Daniel Dolfi\authormark{3} and Aude Martin\authormark{3}}

\address{\authormark{1}Thales SIX, 4 avenue des Louvresses, Gennevilliers, France  \authormark{2}Laboratoire de Physique de l'Ecole Normale Supérieure, 24 rue Lhomond, Paris, France  \authormark{3}Thales Research and Technology, Palaiseau, France }

\email{\authormark{*}bruno.martin@thalesgroup.com} 



\begin{abstract}
Dual-comb LiDARs have the potential to perform high-resolution ranging at high speed. Here, through an implementation involving electro-optic modulators and heterodyne detection, we quantify the ranging systems trade-off between precision and non-ambiguity range (NAR) using a unique performance factor. We highlight the influence of the comb amplitude envelope on the precision with a distance measurement limited by the repetition 
rate of the optical comb. The influence of the combs repetition rate on the NAR and on the precision is illustrated through a setup allowing distance measurement with a tunable NAR. Finally, we demonstrate the impossibility to resolve different targets, quantify the impact on the measured distance and develop on the conditions in which non-linear effects of the interference make the 
measurement impossible. 
\end{abstract}

\section{Introduction}
	Since its beginnings, the application domain of LiDARs (Light Detection and Ranging) has expanded beyond atmospheric and topography measurements to gain focus in surface profiling in industrial environment, 3D mapping for autonomous vehicles or ultra-high precision positioning for satellite formation.
Depending on the targeted application and required performances, many different types of LiDARs were developed \cite{jain2013laser} with various detection schemes (direct or coherent) and types of signals involved (at different wavelengths, pulsed or continuous, etc). Frequency-combs (FC) sources, also widely used in spectroscopy and high-precision metrology, provide an interesting tool for high-resolution and high-speed distance measurements. The resolution of the distance measurement is increased compared to a time-of-flight measurement and the constraints on the speed of the detectors is relaxed \cite{minoshima2000high}. In dual-comb LiDARs, the interference, measured on a detector, between  two frequency combs of slightly different repetition rates allows for a distance measurement with an ambiguity range inversely proportional to the repetition rates of the optical combs and with a sub-millimeter achievable resolution \cite{zhu2018dual}. Particularly, setups based on the coherent detection of an optical comb sent to the target allow for a distance measurement with range and accuracy limited by the frequency comb repetition rate \cite{weimann2015measurement}. In this article this method will be referred to as repetition-rate-limited.	More recently, setups in which the direct detection of two slightly offset frequency combs sent to the target have been proposed \cite{koos2018multiscale,zhang_compact_2015}. This method provides a much larger non-ambiguity range defined by the difference in the two combs repetition rates, allowing the disambiguation of the distance measurement. Other setups involved frequency combs sources as a metrological tool to measure the frequency modulation of a laser used in Frequency Modulated Continuous-Wave LiDARs \cite{coddington2011characterizing} or combined with diffractive optics as a mean to measure distance in different directions simultaneously \cite{riemensberger2020massively}. 
In this article, the critical parameters quantifying the performance of dual-comb LiDAR systems are investigated. We detail in section \ref{sec:repetion_rate_limited} the method used to evaluate the distance as well as the experimental implementation of the repetition-rate-limited setup. It allows us to highlight the direct relationship between the non-ambiguity range and the precision. The influence of the combs properties on the precision are discussed, supported by experimental results. We particularly provide the first investigation on the influence of the combs' amplitude envelopes on the precision. In section \ref{sec:Disambigutaed}, we detail the method allowing a disambiguation of the distance measurement, reaching a tunable NAR up to 375 m. By measuring the precision for different NAR, we illustrate the linear relationship between precision and non-ambiguity range with experimental results. We use this ratio as a performance factor characterizing dual-comb LiDAR systems and provide a novel comparison of previous works found in the literature. Finally, in section \ref{sec:spurious}, we provide the first mathematical description of the effect of a spurious reflection on the synthetic wavelength interferometry technique. This theoretical analysis is backed up by experimental results. 

\section{Repetition-rate-limited distance measurement} \label{sec:repetion_rate_limited}
\begin{figure}
	\centering 
		\includegraphics[width=\textwidth]{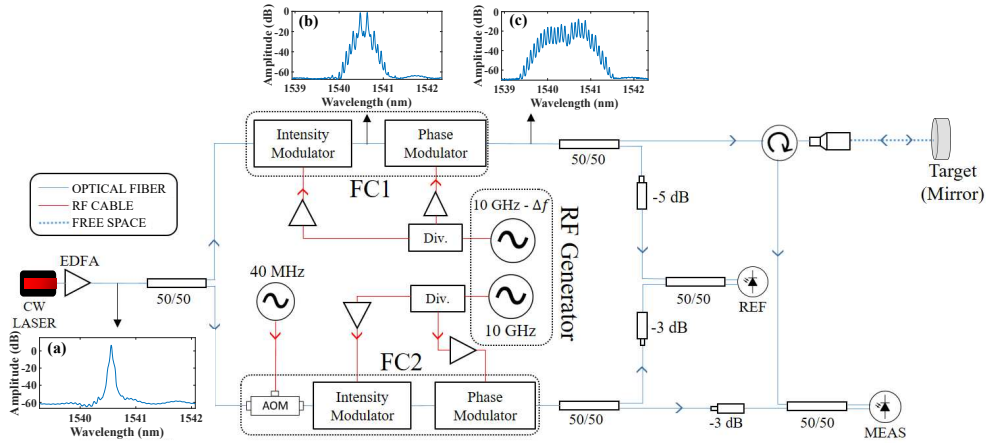}  
		\caption{\label{fig:shortsetup} Experimental setup of the repetition-rate-limited dual-comb distance measurement. The emission of a 1.55 $\mu m$ continuous-wave (CW) laser is split in two in order to generate two optical frequency combs with a $f_{r,s} = 10$ $GHz - \Delta f$ and $f_{r,l} = 10$ $GHz$ repetition rates. The signal optical comb is sent via a collimator to a mirror placed on a linear motorized stage. Both of the combs interfere on reference and measurement balanced photodiodes. EDFA = Erbium-doped fiber amplifier. CW = continuous-wave. RF = Radio-frequency. REF = Reference balanced photodiodes. MEAS = Measurement balanced photodiodes.}
	
\end{figure} 
The repetition-rate-limited dual-comb ranging method relies on two optical combs with slightly different line spacing \cite{coddington2009rapid}. One of them, called the signal comb, is sent through the distance to be measured and is optically sampled by the second optical comb, called the local oscillator. The radio-frequency (RF) comb resulting from this interference is measured on balanced photodiodes both for a measurement and a reference path. The coherence of the optical sampling allows for the amplitude and phase information of the signal comb to be transposed to the RF domain, enabling their measurement and thus the determination of the distance through synthetic wavelength interferometry (SWI) \cite{zhu2018dual}. Previous demonstrations of high-resolution ranging reaching sub-millimeter resolutions and precisions in the order of few micrometers were achieved using this dual-comb technique \cite{Weimann:18, doi:10.1063/1.4999537}. Such setups have also been widely used for spectroscopy \cite{soriano2020common,millot2016frequency}.

The frequency combs are generated from a $\lambda_0 = 1.55~\mu m$ narrow linewidth (4.5 kHz according to the manufacturer) continuous-wave laser (inset (a) in Fig. \ref{fig:shortsetup}), at an optical power tunable between 5 and 15 dBm with an erbium-doped fiber amplifier (EDFA). The optical power is equally split into two paths in order to generate the FC1 and FC2 optical combs by electro-optic modulation \cite{torrescompany2014optical}. In this first implementation, the FC1 comb serves as the signal comb and FC2 as the local oscillator. Each electro-optic modulation comb generation scheme is composed of an intensity modulator followed by a phase modulator (see insets (b) and (c)). The modulators are driven by two RF electrical signals at frequencies of respectively $f_{r,s} = 10 $ $GHz - \Delta f$ and $f_{r,l} = 10$ $GHz$, corresponding to the repetition rates of the signal and the local oscillator optical combs. 
Both RF electrical signals are sinusoidal waves generated from a signal generator whose two outputs are referenced to a shared clock signal. Both of the modulators are driven at their maximum electrical input power. This corresponds to a driving voltage of $1.1$ $V_\pi$ for the intensity modulators and $1.7$ $V_\pi$ for the phase modulators, $V_{\pi}$ being the half-wave voltage. Moreover, the bias voltages of the intensity modulators are adjusted to obtain spectrally flat frequency combs \cite{fujiwara2003optical}. Each of the two optical combs have a total bandwidth of 268 GHz and contain 28 modes, 13 of them being contained in a $\pm$6dB amplitude bandwidth.
We then obtain two combs of optical frequencies : 
\begin{equation}\label{eq:comb} 
	\left \{ 
	\begin{aligned}
		f_{s,m} = f_{0,s} + mf_{r,s} \\ 
		f_{l,m} = f_{0,l} + mf_{r,l} \\ 
	\end{aligned} 
	\right.
\end{equation} 
Where $m$ is the \emph{order} of the frequency modes, and $f_{0,s}$ and $f_{0,l}$ correspond to the central frequencies of the signal and the local oscillator respectively. 
The beating of the two optical frequency combs on the photodiodes generates a comb in the RF domain with a repetition rate of $\Delta f$. Considering the optical frequencies of the two combs and the photodiodes bandwidth (500 MHz), only the beat notes between two frequency modes of the same order can be measured.
Moreover, an acousto-optic modulator (AOM) at a frequency of $f_{AOM} = 40$ $MHz$  is placed before the generation of the local oscillator. This ensure a 40 MHz offset on the RF comb frequencies, preventing aliasing of the negative order modes and reducing the impact of low frequency electrical noises.
The information on the distance is then extracted from the differences between the phases of the interference pattern measured at the reference photodiodes after both combs generation, and the one measured at the measurement photodiodes after the signal comb has been reflected by a mirror at a distance $d$ such as  \cite{weimann2017silicon} :
\begin{equation}\label{eq:distance} 
		d = \Big(\frac{\partial \Delta \Phi_m}{\partial m} (d) - \frac{\partial \Delta \Phi_m}{\partial m}(0) \Big)\alpha^{-1} 
\end{equation}
where $\Delta \Phi_m$ is the difference between the phase of the RF comb mode m on the reference and measurement photodiodes, and
$\alpha = \frac{2\pi f_{r,s}}{c}$.
The two balanced photodiodes voltages are then processed by an analog-to-digital converter (ADC) at a frequency of $f_{ADC} =250$ $MHz$. After a fast Fourier transform (FFT), whose resolution depends on the acquisition time $t_{aq}$, $\frac{\partial \Delta \Phi_m}{\partial m}$ is estimated by linear regression of the measured $\Delta \Phi_m$. 
In order to compensate for the difference of optical distances traveled in the fibers, a first measurement, referred to as a reference position measurement, is needed. This reference position measurement allows for the measurement of $\frac{\partial \Delta \Phi_m}{\partial m}(0)$ requested for the further estimation of the distance, as shown by Eq. \ref{eq:distance}. The target is moved by a distance d, $\frac{\partial \Delta \Phi_m}{\partial m} (d) $ is measured  and the distance difference between the reference position measurement and the second measurement is estimated. We chose to neglect the variation of the distance traveled in fiber between the reference position measurements and the distances estimations since the reference position measurement is repeated before each distance evaluation. Also, the measured distance corresponds to the optical path, thus incorporating the refractive index of air. We also assume that the refractive index of air remains constant during the experiment. These assumptions have been made in regards of the very stable thermal and pressure environment we worked in and the attained precision. Furthermore, dispersive effects are also neglected considering the bandwidth of the combs (few hundreds of GHz). Nevertheless these effects can have a critical influence on the measurements in real conditions \cite{Weimann:18}. \\

\begin{figure}
	\centering
	\includegraphics[width=0.8\textwidth]{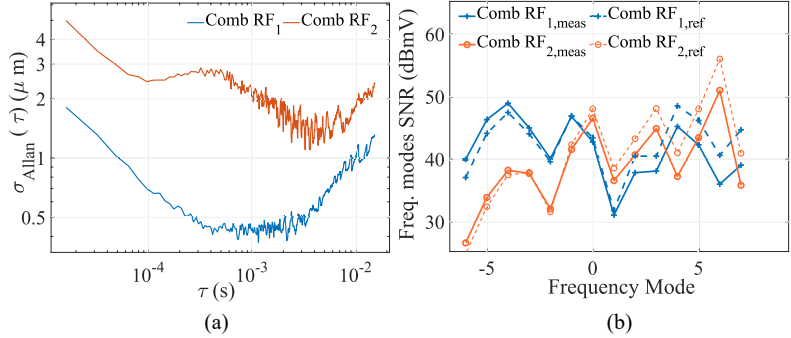} 
	\caption{\label{fig:adev} (a) - Allan deviation for a distance measurement at $d = 5 $ $\mu m$ from the reference position, with a difference in the repetition rates of $\Delta f = 4.4 $ $MHz$ and an acquisition time of $t_{aq} = 16 $ $\mu s$. The same measurement has been done for two different comb amplitude envelopes represented in (b) at the measurement ($RF_{X,meas}$) and reference ($RF_{X,ref}$) photodiodes, for a similar reflected power.}
\end{figure} 

 The reference position is estimated with the ranging of a mirror at a distance of 1 m from the collimator and by averaging the obtained phase difference slopes on 550 time traces of 16 $\mu s$. This increase the precision by averaging most of the zero-mean  high-frequency noises above 115 Hz. This includes the detectors shot noise, thermal electrical noises, and the CW laser relative intensity noise. The influence of these noises on the overall SNR of each frequency mode is more thoroughly described in \cite{Weimann:18}. The mirror has then been translated by a known distance using a motorized translation stage (with a 0.13 $\%$ accuracy according to the manufacturer). As the two measurements are made exactly in the same conditions, we focus our study only on the second measurement. However, it is meaningful to notice that the uncertainty on the reference position measurement will have a similar impact on the accuracy of the result. The reference position and distance measurements are performed in the same conditions and their uncertainties are mainly due to noise. Thus we can approximate that both of the measurements have the same uncertainty, and that the overall uncertainty is equal to the uncertainty of one measurement multiplied by a factor $\sqrt{2}$.

In Fig. \ref{fig:adev}, the Allan deviation of a distance measurement is displayed as a function of the averaging time for an expected distance at 5 $\mu m$ from the reference position and for an integration time of 16 $\mu s$ per measurement. In order to illustrate the influence of the combs' amplitude envelopes, the measurement has been made while applying two different  DC bias on the intensity modulators, providing two different measurement RF combs $RF_1$ and $RF_2$ \cite{fujiwara2003optical}. The two RF combs used for the measurement , $RF_1$ and $RF_2$, each contain 14 modes at the same frequencies. The signal-to-noise ratios (SNR) of each mode are represented in Fig. \ref{fig:adev}.b, both at the measurement ($RF_{X,meas}$) and reference ($RF_{X,ref}$) photodiodes. The reflected optical power is similar for both of the measurements (39 $\mu W$ on the measurement photodiode for comb $RF_1$, 45 $\mu W$ for comb  $RF_2$). The Allan deviation is initially falling as the averaging time increases until it achieves its lowest value of $372$ $nm$ for an averaging time of $1.15$ $ms$ ($1.1$ $\mu m$ at 3.6 ms for comb $RF_2$), coming from the fact that the high frequency noises (shot noise, electrical noises,...) are attenuated through averaging. It then rises again due to the growing impact of low frequency noises (thermal noises). The non-averaged Allan deviation is at $ 1.8$ $\mu m$  for the $RF_1$ comb, and $5.07$ $\mu m$ for the $RF_2$ comb. 

As thoroughly described in the supplementary material, the standard deviation of the slopes of the RF frequency modes phases reference-to-measurement differences, $\sigma_s$, can be measured or estimated knowing the SNR of each frequency mode and  is directly linked to the trade-off between precision ($\sigma_d$) and non-ambiguity range ($NAR$) through:

\begin{align} 
		\sigma_d &= \frac{2 * (NAR)}{\pi} \sigma_s \label{eq:NARstd} \\
		\sigma_s &= \sqrt{\sum_m \frac{c_m^2}{2}\big(\ \frac{1}{SNR_{m,meas}} + \frac{1}{SNR_{m,ref}}\big)} \label{eq:sigmas}
\end{align}

with $c_m$ the linear regression coefficients more thoroughly detailed in the supplementary material, and $SNR_{m,meas}$, $SNR_{m,ref}$ the signal-to-noise ratios of the frequency modes $m$ of the measurement and reference RF combs respectively. By computing the expected $\sigma_s$ through Eq. \ref{eq:sigmas} with the results from Fig.\ref{fig:adev}.b we find an expected $\sigma_s$ of $0.6$ $mrad$ for comb $RF_1$ and $1.7$ $mrad$ for comb $RF_2$. Considering the NAR of the system and Eq. \ref{eq:NARstd}, it accounts to an estimated measurement standard deviation of $\sigma_d = 2.9$ $\mu m$ for comb $RF_1$ and $\sigma_d = 8.1$ $\mu m$ for comb $RF_2$. By generating, in the same conditions, an optical signal comb with a higher repetition rate, the measurement will achieve a greater precision and a shorter NAR. However, as long as the number of comb lines and their amplitude envelope stay unchanged, the ratio between $\sigma_d$ and the NAR will stays the same and is quantified by $\sigma_s$. 
	
The estimated results of $\sigma_d$ are consistent with the non-averaged precisions presented in Fig. \ref{fig:adev}.a and illustrate the influence of the RF comb amplitude envelope on the measurement precision. Indeed, the values of the linear regression coefficients $c_m$ imply that the SNR of higher frequency modes (on the sides of the comb) have a larger impact on the overall precision. It explains the fact that, for the same detected power, the results obtained with the comb $RF_1$ favoring the SNR of high orders modes provide a better overall precision than the results obtained with the comb $RF_2$, favoring the SNR of low order modes (in the center of the comb). We attribute the difference between the estimation of $\sigma_d$ and the measurements to the uncertainty of the SNR measurement. $\sigma_s$ therefore constitutes a unique performance index describing the performance of the ranging system. It takes into account both the precision and the non-ambiguity range, and depends only on the comb amplitude envelope and on the detection SNR. This trade-off between non-ambiguity range and precision is tuned by adjusting the repetition rate of the signal comb for a given comb generation method. 
\begin{figure}[!b]
\centering
\includegraphics[width=\textwidth]{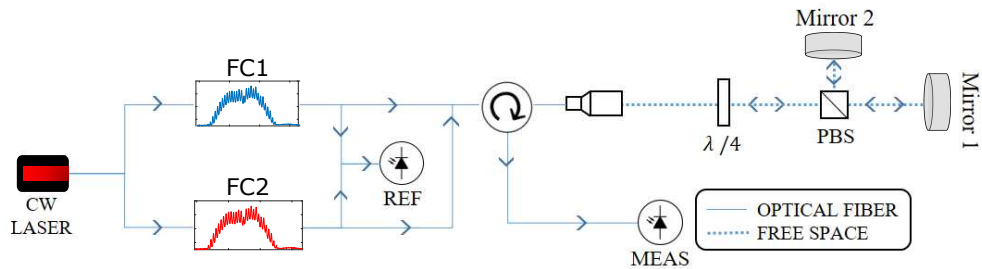} 
\caption{\label{fig:longrange} Simplified scheme of the disambiguated distance measurement setup. Both of the optical combs are generated through the same electro-optic modulation scheme as depicted in Fig \ref{fig:shortsetup}. On one hand, the two optical combs FC1 and FC2 interfere at the reference photodiode, just after their generation. On the other hand, the two combs are combined before the circulator, and sent to the target. Mirror 1 is placed on a linear motorized translation stage, whereas mirror 2 is fixed at the other side of the room. FC : Frequency comb. REF : Reference photodiode. MEAS : Measurement photodiode. $\lambda / 4$ : Quarter-wave plate. PBS = Polarizing beamsplitter. }

\end{figure}
\section{\label{sec:Disambigutaed} Disambiguated distance measurement using the interference of two optical frequency combs}

\begin{figure}[!t]
	\centering
	\includegraphics[width=12cm]{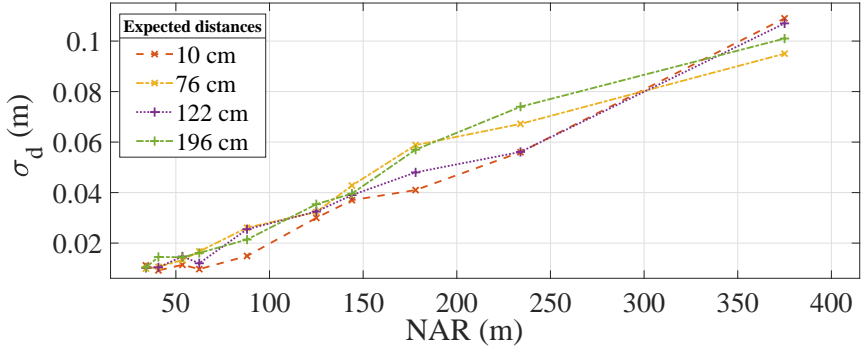} 
	\caption{\label{fig:narstd} Standard deviation of the disambiguated distance measurement as a function of the non-ambiguity range (NAR). Each line corresponds to a same expected distance, measured multiple times while changing the NAR of the system between each acquisition. }
	
\end{figure}

In the previous method, a main limitation is the non-ambiguity range, that is limited by the line spacing of the signal comb. A disambiguation method was proposed \cite{koos2018multiscale} where both combs are sent to the target. This specific method has, in our knowledge, never been demonstrated experimentally. As explained in the simplified experimental setup in Fig. \ref{fig:longrange}, both of the combs (FC1 and FC2) are coupled right before being sent to the target. This change in the repetition rate of the signal comb has for consequences to change the $\alpha$ factor from $\frac{2 \pi f_{r,s}}{c}$ to $\frac{2 \pi \Delta f}{c}$. Thus, the resulting measurement will have a greater NAR, at the price of a decreased precision. Further increase of the NAR requires smaller $\Delta f$ and therefore a greater acquisition time in order to be able to resolve each RF comb line. In our setup, the NAR is easily tunable by changing the repetition-rate difference between the two combs, $\Delta f$, from $200$ $kHz$ to $2.2$ $MHz$. It accounts for a theoretical NAR varying from 34 m to 375 m. The RF combs contain 13 frequency modes whose intensity distribution follows the one obtained in Fig. \ref{fig:adev}.b. For each NAR, we measure, for several expected distances, the standard deviation of 550 distance measurements with an integration time varying between 65 and 262 $\mu s$.
The precision of our results obtained with the disambiguated setup is presented in Fig. \ref{fig:narstd}. Experimentally, we retrieve the linear relationship between the NAR and the precision found in Eq. \ref{eq:NARstd}. The slope of the results obtained in Fig. \ref{fig:narstd} correspond to $\sigma_s$ and quantitatively describe the performance of the setup in term of precision and non-ambiguity range. By linear regression analysis and according to Eq. \ref{eq:NARstd}, it is therefore possible to determine more accurately the coefficient $\sigma_s$. We obtain a result of $\sigma_s = 0.437$  $mrad$ ($R^2 = 0.9806$).

%

\begin{figure}[!t]
	\centering
	\includegraphics[width=9cm]{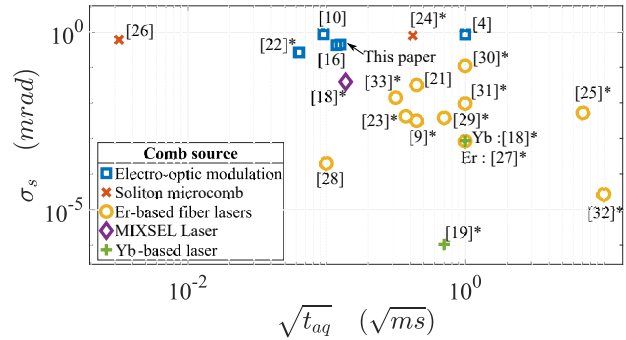} 
	\caption{\label{fig:comp} Ratios between the non-ambiguity range and the precision, quantified by $\sigma_s$, as a function of the square root of the acquisition time $\sqrt{t_{aq}}$ for the main results on dual-comb based distance measurement found in the literature. The comb source used for each result is specified. Results labeled with an asterisk are based on a time-of-flight measurement }
\end{figure}

Considering the influence of the acquisition time on the signal-to-noise ratio \cite{coddington2016dual}, the frequency domain resolution and thus on the precision of the measurement , we choose $\sigma_s (\sqrt{t_{aq}})$ as the main parameter describing the performance of dual-comb distance measurement and compare in Fig. \ref{fig:comp} our results to those already found in previous works. It updates previous comparisons \cite{Weimann:18} by taking into account the trade-off between precision and non-ambiguity range. As the comb source properties have a strong influence on $\sigma_s$, the comb generation technique used in each work is specified. Every result has been obtained in an indoor environment with small temperature and pressure variations. Highly compelling works in which long free-space propagation distances influence the precision through atmospheric effects are presented in the literature \cite{doi:10.1063/1.4999537}, but not included here to avoid unfair comparison.

  These results reflect that better performances can be achieved by generating broader and flatter combs. Systems using Er-doped fiber lasers or Yb-based lasers demonstrate a better coefficient $\sigma_s$ as they emit stable and flat combs, but their low repetition rates ($\sim 100 MHz$) increase the required acquisition time. On the other hand, the recent development of integrated photonics and more specifically of microresonators offers soliton microcombs with a high repetition rate ($\sim$ 100 GHz), enabling high precision results with very short acquisition times. Similarly, interesting results have been obtained with a mode-locked integrated external cavity surface-emitting laser (MIXSEL) in \cite{nurnberg2021dual} at a repetition rate around $2$ $GHz$, undermined by the laser low output power. EOM-based system provide cost-effective stable combs with a high repetition rate ($\sim 10 GHz$), allowing precise measurement at high speed. More crucially, the high interest of EOM-generated combs lies in their repetition rate tunability. More generally, we note that even though the comb generation technique highly influence the results regarding the comb frequency span, stability and intensity distribution, tremendous results have been obtained through original NAR disambiguation techniques. For instance, results presented in \cite{fellinger2021simple} rely on the modulation of the optical-comb repetition frequency to produce sidebands creating very large synthetic wavelengths. It allows additional measurements of the distance up to potentially hundreds of kilometers while preserving the precision offered by conventional dual-comb LiDARs. Results labeled as time-of-flight (TOF) measurement consist in measuring the time difference, or in a similar way the phase shift, between the reference and measurement pulses \cite{zhu2018synthetic}. Since TOF measurements require to resolve both pulses, their resolution is limited by the sampling frequency. As it nevertheless leads to performance factors $\sigma_s$ similar to the one obtained with synthetic wavelength interferometry, we included TOF results in Fig. \ref{fig:comp}. Furthermore, the performance of setups in which the distances are estimated through the evaluation of the pulses temporal position does not depends on $\sigma_s$ as the measurement does not depends on phases difference estimation. It is nevertheless relevant to compare their performance in terms of non-ambiguity range to precision ratio. Therefore, we included them in the Fig. \ref{fig:comp} and their 'relative' $\sigma_s$ was computed from the ratio between their precision and their non-ambiguity range.



\nocite{weimann2015measurement}
\nocite{weimann2017silicon} 
\nocite{lee2013absolute} 
\nocite{Teleanu:17} 
\nocite{liu2011sub} 
\nocite{suh2018soliton} 
\nocite{hu2021dual}
\nocite{trocha2018ultrafast} 
\nocite{nurnberg2021dual}
\nocite{fellinger2021simple}
\nocite{zhou2020multi}
\nocite{li2020high}
\nocite{zhang2014absolute}
\nocite{lin2017dual}
\nocite{jiang2021aliasing}
\nocite{zhu2019two}
\nocite{wright2021two} 
\section{Distance measurement of multiple targets} \label{sec:spurious}

When sensing real environments, the backscattered signal might be composed of numerous successive reflections coming from different targets, each having its respective distance, reflectivity and velocity. Even if only one target is aimed at, disturbances placed in the light path can create spurious reflections. The ability to differentiate multiple targets is therefore a key feature of LiDAR systems.\\
\begin{figure}
	\centering 
	\includegraphics[width=\columnwidth]{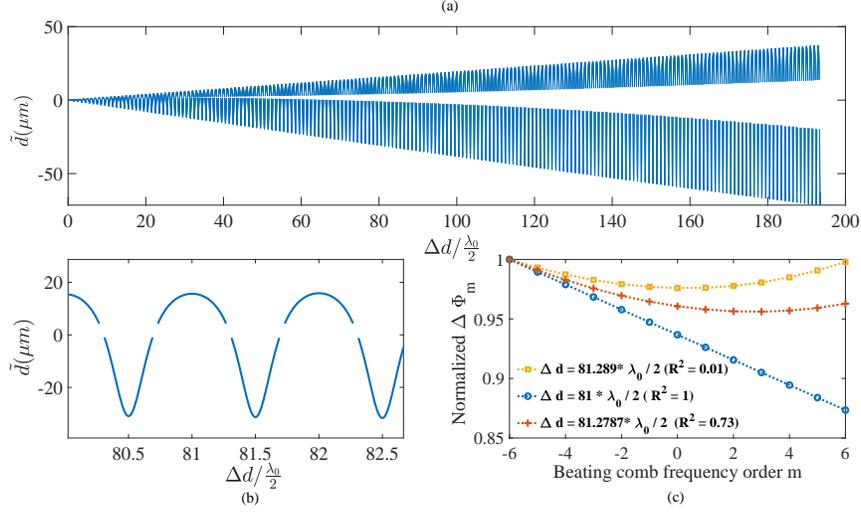} 
	\caption{\label{fig:spurious_reflection} (a) - Simulation of the error on the measured distance created by a spurious reflections $\tilde{d}$ as a function of the distance of the spurious reflection from the main distance $\Delta d$ (normalized by $\frac{\lambda_0}{2}$) with an initial distance $d_1 = 2$ $\mu m$, a 13 teeth symmetrical signal optical comb, and a relative reflected signal amplitude ratio $p = 0.25$. Only the measures for which the coefficient of determination $R^2$ is higher that 0.9 are plotted. (b) - Inset of the simulation plotted in (a). (c) - Normalized phase differences $\Delta \Phi_m$ of the comb teeth for 3 values of $\Delta d$ corresponding to the simulation plotted in (a) and (b). The data corresponding to the red and yellow curves are not plotted in (a) and (b) since their coefficient of determination is lower than 0.9.}
\end{figure}
In the case where backscattered signals coming from multiple targets are hitting the photodiodes during the integration time $t_{aq}$, the time-frequency analysis is made on the interference of the multiple reflections. \\ 
If the targets do not have the same speed, each of them will produce shifted RF frequency combs due to the different Doppler frequency shifts. This allows for a differentiation of each target as long as the differences in the Doppler shifts are higher than the fast Fourier transform (FFT) resolution bandwidth and lower than half of the RF comb repetition rate. If, however, the targets have the same velocity, they will produce the same RF combs frequencies. In this case, the measured phase of each frequency mode will result as a weighted sum of the phases corresponding to the reflection of each target. Though this issue has already been mentioned and tackled through time-gating \cite{coddington2009rapid}, it has in our knowledge never been properly quantified.\\
In the case of two static targets at distances $d_1$ and $d_2$, whose reflections of amplitude $A_1$ and $A_2$ are integrated during $t_{aq}$, we define $\epsilon_{\psi,m}$ as the phase shift on the frequency mode m induced by the second reflection compared to the one obtained for a single target (more details are provided in the supplementary material). We find : 
\begin{equation}\label{eq:epsilonphi} 
		\epsilon_{\psi,m} = arctan\bigg(\frac{(1-p)sin(\frac{4\pi d_1 f_{s,m}}{c}) + p \, sin(\frac{4\pi d_2 f_{s,m}}{c})}{(1-p)cos(\frac{4\pi d_1 f_{s,m}}{c}) 
			+ p \, cos(\frac{4\pi d_2 f_{s,m}}{c})}\bigg) \\ 
		- \frac{4\pi d_1f_{s,m}}{c}
\end{equation}
Where $p = \frac{A_2}{A_1 + A_2}$ depends on the ratio of the reflections amplitudes of the two targets. Knowing the shift on each frequency mode phase, we derive by linear regression analysis and according to Eq. \ref{eq:distance} the error $\tilde{d}$ on the distance information  $d_1$ induced by a spurious reflection at a distance $d_2$, described as: 
\begin{equation}\label{eq:dtilde} 
		\tilde{d} = \frac{c\sum_m \big(\epsilon_{\psi,m}c_m\big)}{4 \pi f_{r,s}}	
\end{equation}

 The simulated results of Eq. \ref{eq:dtilde} as a function of the distance difference $\Delta d = d_2 - d_1$ between the two targets, relative to half of the optical carrier wavelength $\lambda_0$ are plotted in Fig. \ref{fig:spurious_reflection}.a.
The error on the determination of the distance increases as the distance between both targets increases with a periodicity of $\lambda_0/2$ as described in Eq. \ref{eq:epsilonphi}. Thus $\tilde{d}(\Delta d)$ is composed of a periodic function of period $ \frac{\lambda_0}{2}$ multiplied by an envelope depending on $p$. For higher values of $p$ (\textit{i.e} for a spurious reflection with a greater backreflected power) the induced shift $\tilde{d}$ in the estimated distance information will be greater even in a situation where the targets have the same distance difference $\Delta d$.
The error varies continuously except for a set of parameters leading to a misevaluation in the linear regression of $\frac{\partial \Delta \Phi_m}{\partial m}$ ($R^2$ < 0.9). These specific cases correspond to the distances not plotted in Fig. \ref{fig:spurious_reflection}.a and \ref{fig:spurious_reflection}.b. Indeed, perturbations caused by multiple reflections will change the relation between $\Delta \Phi_m$ and the beating mode orders $m$ and can lead to a non-linear relation as evidenced in Fig. \ref{fig:spurious_reflection}.c. As each phases difference $\Delta \Phi_m$ is shifted by $\epsilon_{\psi,m}$, the loss of linearity of $\Delta \Phi_m (m)$ comes from the fact that $\frac{\partial ^2 \epsilon_{\psi,m}}{\partial m^2} \neq 0$. By analyzing $\frac{\partial^2 \epsilon_{\psi,m}}{\partial m^2}$, we determine that $\Delta \Phi_m (m)$ stays approximately linear only for small values of $\Delta d$ close to a multiple of $\frac{\lambda_0}{2}$ (see supplementary material). This issue is illustrated in \ref{fig:spurious_reflection}.c  by the simulated values of $\Delta \Phi_m (m)$ for $\Delta d = 81 * \frac{\lambda_0}{2}$. While, for this value, the coefficient of determination indicates the goodness of the fit ($R^2 = 1$), the measurement of the distance is inaccurate and corresponds to a weighted sum of the two distances, determined by Eq. \ref{eq:dtilde}. For other values, such as the two other curves presented in Fig. \ref{fig:spurious_reflection}.c, the linearity is broken ($R^2 << 0.9$), making the fit impossible. The corresponding estimated distances are thus not plotted in Fig. \ref{fig:spurious_reflection}.b. While the loss of linearity of $\Delta \Phi_m (m)$ make the spurious reflection easily detectable in the latter case, in the first case errors of tens of microns for an expected main distance $d_1 = 2$ $\mu m $ arise as shown in Fig. \ref{fig:spurious_reflection}.a, without any mean of detecting it. \\  

\begin{figure}
	\centering
	\includegraphics[width=12cm]{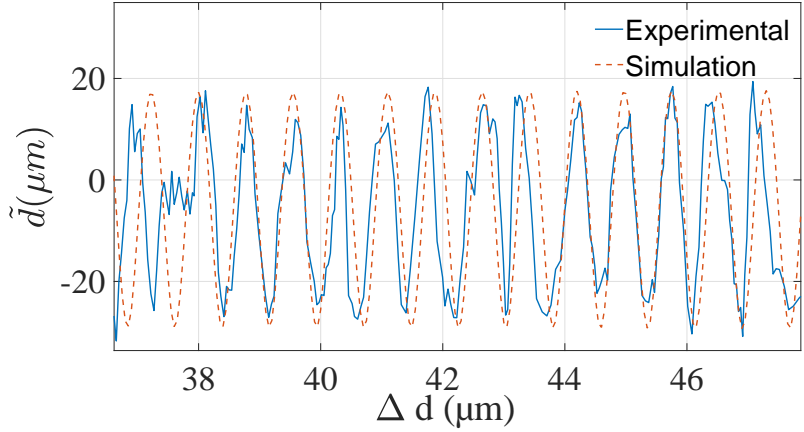}
	\caption{Measured distance shift $\tilde{d}$ coming from the interference of the backreflected signal of mirror 1 and mirror 2 in Fig. \ref{fig:longrange} as a function of  the distance of the spurious reflection from the main distance $\Delta d$. The solid blue line correspond to the experimental results, and the dashed orange line to the model developed in the text.}
	\label{fig:vibrometry10both}
\end{figure}

Our model was verified experimentally by using a $\frac{\lambda}{4}$ waveplate to change the power ratio between the signal reflected by two mirrors M1 and M2 as illustrated in Fig. \ref{fig:longrange}. This way, one of the mirror acts as the main target at distance $d_1$, and the second one as the spurious reflection at distance $d_2$, with a relative amplitude $p$ tuned by rotating the waveplate. Mirror 1 was then translated at a speed $v_t$ in order to modify $\Delta d$. The experimental and simulated results are obtained for $p\simeq 0.06$, $v_t = 1$ $\mu m/s$. The RF combs have 14 modes and a repetition rate of $\Delta f = 6.1$ $MHz$. The results are compared to the simulation in Fig. \ref{fig:vibrometry10both}. We can see a reasonable correlation between our model and the experimental results. The phase shifts in the experimental results are attributed to non-linearities in the translated stage velocity. 
We also notice the $0.7 $ $\mu m \simeq \frac{\lambda}{2}$ periodicity of $\tilde{d}$ $(\Delta d)$.

Multiple reflections will therefore either false the distance evaluation or make the measurement impossible due to the loss of linearity of $\Delta \Phi_m$. The single extracted distance makes multiple targets indistinguishable and the inaccuracy on the main distance information in the case of spurious reflections is far from negligible even with spurious reflections at low relative amplitude. \\   



\section{Conclusion}

Through the analysis of a conventional dual-comb LiDAR technique based on synthetic wavelength interferometry, we were able to understand the influence of the comb generation parameters on the overall precision of the measurement. Particularly, the first quantitative analysis regarding the influence of the dual-comb interferogram spectrum amplitude envelope on the precision has been presented and illustrated experimentally.

Moreover, through the first experimental demonstration of the disambiguated setup presented in \cite{koos2018multiscale}, we were able to experimentally demonstrate the linear relationship between non-ambiguity range and precision. This trade-off is quantified by the standard deviation of the phases differences slopes , $\sigma_s$, depending largely on the amplitude envelope of the frequency combs. 

This leads to a single performance index illustrated through the comparison of results found in previous related works. It allows to better take into account the non-ambiguity range in precisions results, as these last two are closely linked. Furthermore, it allows to better predict the performance of a system knowing the amplitude envelopes and repetition rates of the generated combs as well as the sources and detectors specifications. Experimentally, a performance factor of $\sigma_s = 0.437$ $mrad$ has been obtained for an integration time of $16 $ $\mu s$. It accounts for a precision to non-ambiguity range ratio of $2.8.10^{-4}$.\\ 
Finally we demonstrated through a novel mathematical analysis the inability of the architecture to resolve multiple targets, unless they move at different velocities and if the acquisition time is high enough to provide a frequency resolution enabling to distinguish the targets frequency modes shifted by Doppler effect. This is a critical issue regarding uncontrolled environment utilization that could hinder its use as a LiDAR. Nevertheless, this impediment can be solved by combining this architecture with other techniques such as time-of-flight measurement. \\
Lately, recent progress in integrated photonics have enabled the demonstration of flat and broad combs with increased output power and will encourage the use of dual comb ranging as a mean to increase the resolution and acquisition speed of ranging systems. Furthermore, original methods such as repetition rate modulation \cite{fellinger2021simple} or two-color dual-comb ranging \cite{zhu2019two,lin2017dual} provide remarkable improvements of the non-ambiguity range to precision ratio without the need of a complex comb generation technique.
In the future, investigation of the effects of the free-space propagation such as atmospheric perturbation will be carried out. It will allows to take into account the influence of the measurement environment. Apart from LiDAR applications, the analysis presented in this work can be applied to other techniques relying on the estimation of the phases change as a function of the frequency. For instance, recent results in dual-comb holography \cite{vicentini2021dual} rely on the use of multiple comb lines to perform holographic measurement by unwrapping the phases differences of multiple synthetic wavelengths. Therefore, the trade-off between non-ambiguity range and precision is similar to the one presented in this article. It reflects that the study developed in this paper can be appropriate for a broad applications spectrum.
\section*{Funding}
The authors acknowledge financial support by the Qombs Project [FET Flagship on Quantum Technologies grant No 820419] 	“Quantum simulation and entanglement engineering in quantum cascade laser frequency combs”.
\section*{Disclosures}
The authors declare that there are no conflicts of interest related to this article.
\section*{Data availability}
Data underlying the results presented in this paper are not publicly available at this time but may be obtained from the authors upon reasonable request.\\ \\ 
See Supplement 1 for supporting content
\bibliography{lidar}






\end{document}